# Active optics and coronography with the Hubble Space Telescope

Fabien Malbet, Michael Shao and Jeffrey Yu

Jet Propulsion Laboratory, Spatial Interferometry Group
4800 Oak Grove Dr., M.S. 306-388, Pasadena, CA 91101

## ABSTRACT

In the field of planet and proto-planetary disk detection, achieving high angular resolution and high dynamic range is a necessity. Ground images are blurred by atmospheric turbulence and therefore adaptive optics is necessary to achieve sub-arcsecond images. Coronography coupled with adaptive optics on ground-based telescopes is a way to get both good spatial resolution and high dynamic range. However because of the atmospheric turbulence, ground based systems are still not capable of detecting planets. The Hubble Space Telescope (HST), due to its location above the atmosphere, would be capable if the mirrors were perfect. However, because of the residual figure errors on the primary and on the secondary, HST has a scattered light level that prevents it from detecting extra-solar planets.

Our simulations show that by using an active mirror (400-1000 actuators) in the optical path of HST with adaptive optics, we can correct the mirror errors and decrease the scattering level by a factor $10^2$ (from $10^{-4}$ to $10^{-6}$ fainter than the star). Furthermore, by controlling the spatial frequencies of the active mirror with a *dark hole* algorithm we can decrease the scattering level in image zones where planet detection is likely. Using this technique, we have succeeded in decreasing the scattering level to $3 \times 10^{-8}$ of the star intensity within 1 arcsec from the central star. This will allow the detection of a Jupiter-like planet $10^{-9}$ times dimmer than the central star located 10 pc away in 1 hour of integration time with a signal-to-noise ratio of 5.

This paper describes the method used to determine the actuator strokes applied to a deformable mirror to achieve planet detection and the design of a coronograph which implements this novel technique.

## 1. INTRODUCTION

Since the discovery of the planet Pluto, the $9^{th}$ planet of the solar system, many attempts have been made to detect planets outside of our planetary system. This includes both indirect techniques such as astrometry, radial-velocity measurements, photometry and direct techniques such as coronographic imaging.[1]

Because of its location and large collection aperture, HST is a good candidate for direct imaging of planets. Except for the large amount of spherical aberration, HST has an extremely precise primary mirror ($\lambda/60$ at $0.8\mu m$). However the Space Telescope is not expected to be able to detect planets, even with the new corrective optics (COSTAR and WF/PC2), because the high spatial frequency figure errors on the primary mirror are too large. These errors will generate scatter from the star and prevent planet detection, even with the use of a coronograph. Therefore we propose to use active optics combined with a coronograph to increase the detection capability of the Space Telescope.

All the intensity measurements calculated in this paper are referenced to the star intensity at the center of the telescope PSF before any instrument which modifies the star light distribution (e.g. coronograph, active mirror).

First we will review the major obstacles to detecting planets with the Space Telescope (§2.). An overview (§3.) of the active optics technology needed is then presented. The following section (§4.) will describe how we use the active optics to get a low scattering level and then we will describe the coronographic lay-out (§5.).

## 2. PLANET DETECTION IMAGING ISSUES

The key problem in planet detection is that the planet is about $10^{-9}$ times fainter than the star. Consider a planet located at 0.5 arcsec which corresponds to 5 AU, the Jupiter-Sun distance, at 10 pc. Figure 1 shows the signal-to-noise ratio (SNR) for the case of the Hubble Space Telescope. We can see that in order to use reasonable exposure time, we need a background level $10^7$ times lower than the star radiation. There are two origins for the background: the diffraction wings from the star (cf. § 2.1.) and the scattered light (cf. § 2.2.).

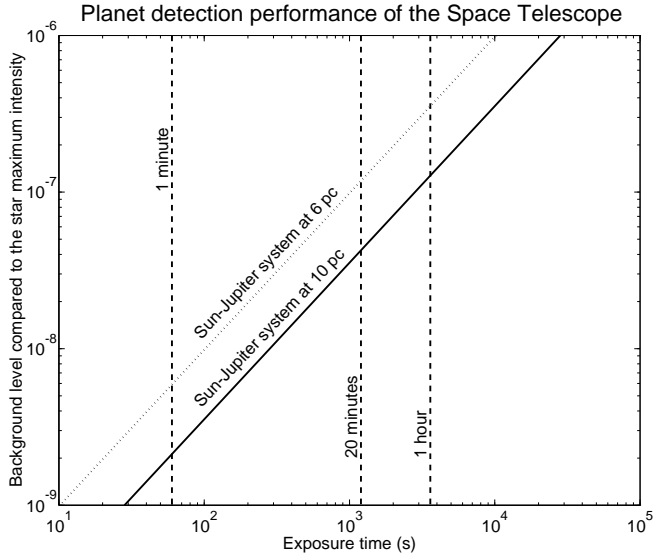

| Parameters | |
|---|---|
| Total transmission | 46% |
| Collecting Area | 4 m$^2$ |
| Central Wavelength | 0.8 $\mu$m |
| Bandwith | 0.4 $\mu$m |
| Star | Sun |
| Planet | Jupiter |
| **Flux at 10 pc:** | |
| • star | $8 \times 10^8$ photons/s |
| • planet | 0.8 photons/s |
| **Flux at 6 pc:** | |
| • star | $2 \times 10^9$ photons/s |
| • planet | 2 photons/s |

**Figure 1:** SNR calculation in the case of the Hubble Space Telescope and for a Sun-Jupiter system at 6 and 10 pc.

### 2.1. Diffraction

Consider the case where the telescope is perfect and nothing perturbs the incoming stellar wavefront. Let us also assume that both the star and the planet are not resolved by the telescope. Because of the limited size of the telescope, the stellar beam will be diffracted as an Airy pattern and the planet buried in the point spread function of the star (see fig. 2). If $\pi(u,v)$ is the pupil function then the intensity of the corresponding point spread function (PSF) is given by

$$I_0 = \left| \widehat{\pi(u,v)} \right|^2 , \qquad (1)$$

where the symbol "$\widehat{\phantom{x}}$" is the Fourier transform operator.

The diffracted light contributes to the photon background and actually prevents any detection of planets. In the case of HST, the diffracted light level at the location of the planet is about $10^{-4}$ times smaller than the central intensity of the star. That means that the background level would be $2 \times 10^5$ photons/s compared to the 2 photons/s coming from the planet located at 6 pc (see Table in Fig. 1). The SNR in one hour of integrating time would only be 0.3, which means that detecting a planet would require an exceedingly long integration time.

### 2.2. Scattering

Let us now consider the case of an actual mirror. The HST mirror has a very accurate surface, but there are some residual errors on the surface. **Scattering** will be defined as any diffraction pattern created only by the small random figure aberrations on the mirror surface, i.e. by $\phi_1(u,v)$, the phase error function. Therefore, if the mirror was perfect, $\phi_1(u,v)$ would be equal to zero at every point of the pupil. The intensity of the actual PSF is then

$$I_1 = \left| \widehat{\pi(u,v) e^{i\phi_1(u,v)}} \right|^2 . \qquad (2)$$

In order to separate the effect of diffraction from the effect of scattering, we will focus on the difference between the actual PSF and the perfect point spread function (generated by a perfect wavefront, cf. § 2.1.). We perform this difference with the complex amplitude in the pupil plane. This is similar to using a Michelson interferometer in the pupil plane with a differential optical path length equal to half a wavelength.

Therefore the scattering is expressed as:

$$S_1 = \left[ \widehat{\pi(u,v) \left( e^{i\phi_1(u,v)} - 1 \right)} \right] \qquad (3)$$

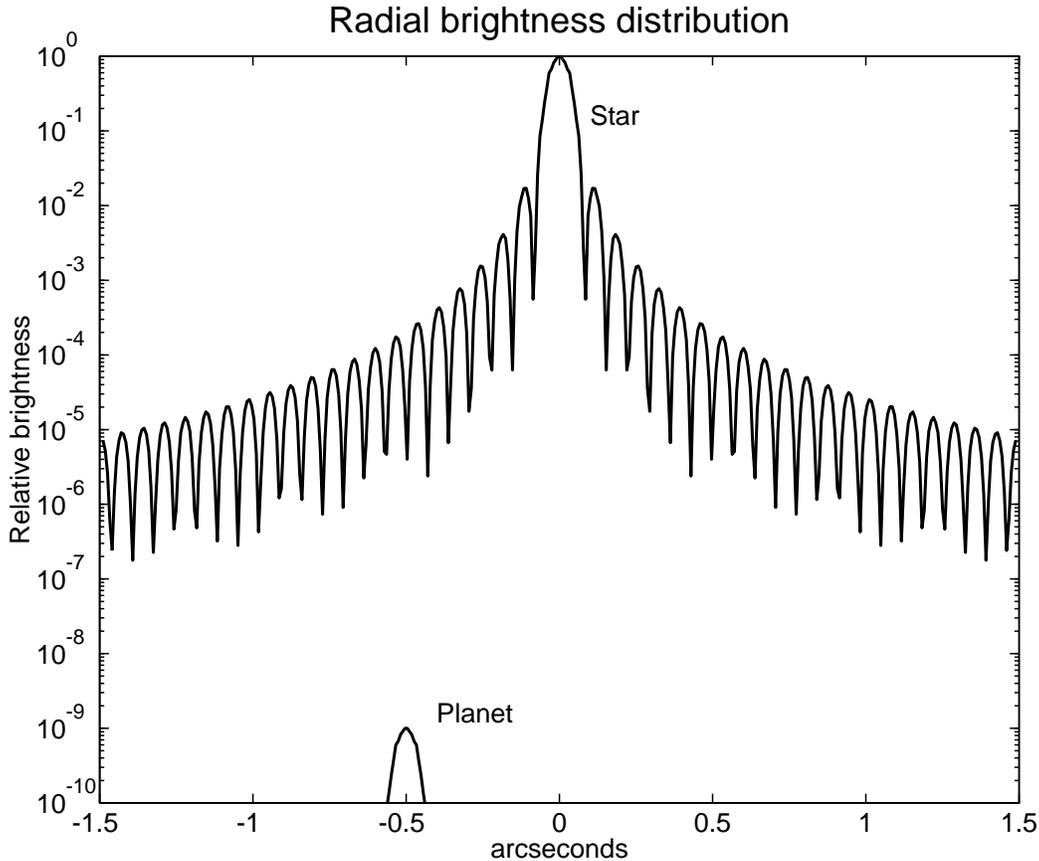

**Figure 2:** Radial brightness of the star. The planet is located a 1 AU and is $10^{-9}$ times fainter than the star.

If the phase errors are small enough, as in the case of HST, then the scattering can be approximated by:

$$S_1 \approx I_0 * \left| \widehat{\phi_1(u,v)} \right|^2 , \qquad (4)$$

where $I_0$ is the previously defined PSF intensity distribution.

The image of the phase errors of the Hubble Space Telescope is shown on Fig. 3. The data were taken from a report by Roddier & Roddier[2] where phase-retrieval techniques were used on 8-9 WF/PC-1 images during the HARP (Hubble Aberration Retrieval Program) program to assess the spherical aberration on HST.

The resulting scattering is shown on Fig. 4. We can see that the scattering level is only about $10^{-5}$ to $10^{-4}$ times dimmer than the star over the whole image and will prevent the detection of a planet. With a coronograph we can reduce the main diffraction effect, but the background will still be limited by this scattering. The SNR results are comparable to those obtained in § 2.1., i.e. the needed integration time to obtain a good SNR is too long.

The diffracted and scattered lights are two additive phenomena. We solved the two problems separately, the first with a coronograph (§ 5.) and the second with an active mirror (§ 4.). In this paper, we propose to use a deformable mirror to lower the scattering level in part of the image and to increase the detection performance in a region where the planet is expected.

## 3. SPACE ACTIVE OPTICS

The space based coronograph will require a high density deformable mirror with low stroke to correct for wavefront errors at low spatial scales. Because of packaging constraints, the image of the pupil on the deformable mirror will not be very large. As an example, a design for the HST advanced camera[3] resulted in a 2.5 cm diameter for the re-imaged pupil. With actuator spacings of 1 mm, this would result in 490 actuators across the mirror, and then the deformable mirror would be capable of only correcting errors in the HST primary at spatial scales larger than 9.6cm.

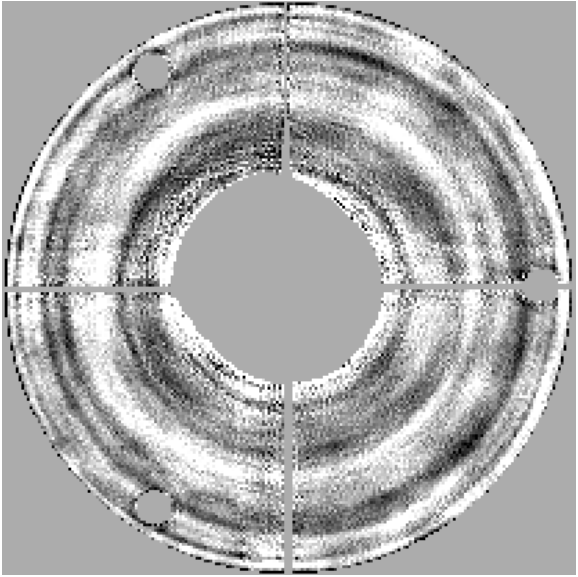
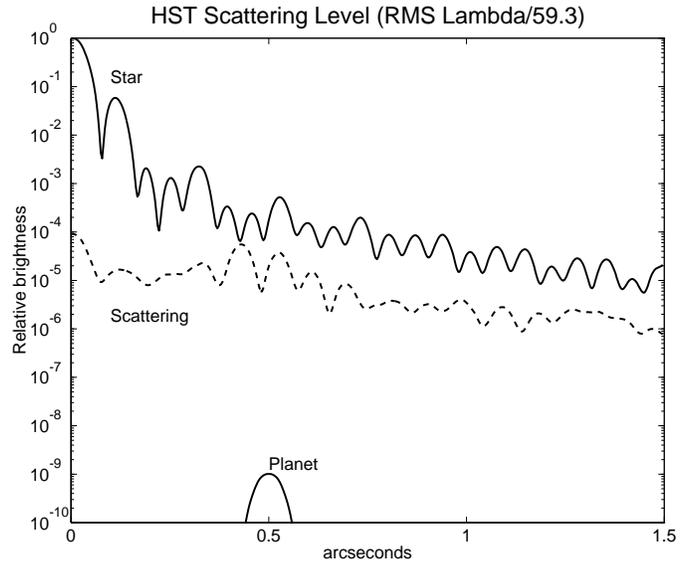

**Figure 3:** HST phase map

**Figure 4:** HST scattering level

**Table 1:** Space active optics vs. ground adaptive optics specifications

| Parameter | Ground (D=10m, $r_0$=20cm) | Space |
|---|---|---|
| # actuators | 2500 | 300-1200 |
| actuator stroke | 10$\mu$m | 50 nm |
| actuator resolution | 10 nm | 0.1 nm |
| servo bandwidth | 50-200 Hz | <0.001 Hz |
| wavefront sensor | Hartman/curvature etc. | science camera (using phase retrieval) |

### 3.1. Space Adaptive Optics vs. Ground Adaptive Optics

While the adaptive optics (AO) concept is the same for ground and for space, its implementation in space and on the ground are very different. Ground-based AO systems are designed to remove the time-varying phase errors introduced by atmospheric turbulence. Space AO has an entirely different goal, namely, to correct the fabrication errors in the large light-collecting optics and to thereby reduce the scattered light caused by those imperfections. Table 1 shows the relevant parameters of space vs. ground AO systems.

Instead of correcting the wavefront for turbulence 100 times a second, space AO is a one-time-only (or e.g. daily update) procedure. In space there is no need for a separate wavefront sensor; the science camera can be used to detect the wavefront error. Phase retrieval techniques are used on the science data to determine mirror correction. Perhaps the biggest difference is in the active mirror. On the ground the active mirror must remove the effects of turbulence which can require up to 10 $\mu$m of optical path for an 8-10 m telescope. In space because the primary and secondary mirror is well polished, the required stroke is very small, 50 nm at most. This can be implemented with a new generation of active mirrors for space which can be built at low cost.

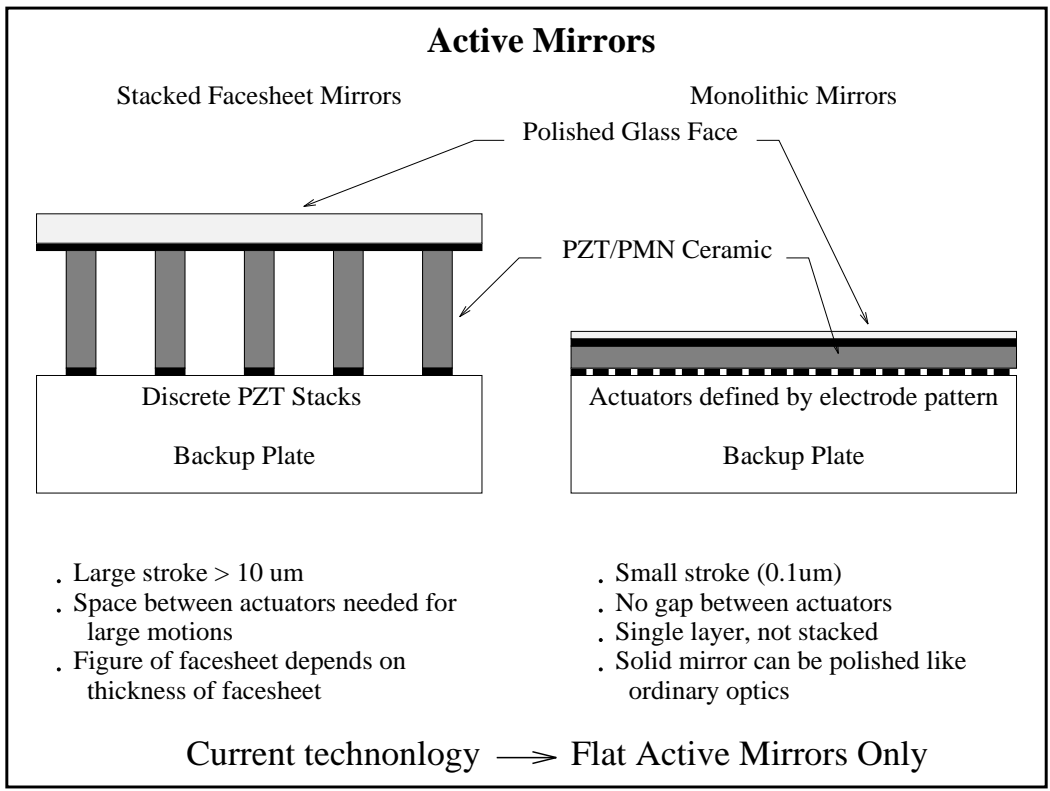

**Figure 5:** Comparison between stacked facesheet mirror and monolithic mirror

### 3.2. Integrated/Monolithic Active Mirrors

Active mirrors for ground-based telescopes come in several flavors. Some are based on stacks of PMN/PZT's with face sheets. Others are based on PZT bimorphs. Yet others are based on PZT/PMN tubes. The earliest deformable mirror was based on a monolithic piece of PZT with an electrode pattern deposited on the PZT to define the actuators. This type of active mirror is no longer used because the stroke of a single PZT layer is insufficient to correct the 10 $\mu$m effects of atmospheric turbulence. For space applications, however, this single layer of PMN/PZT is sufficient to change the optical path by 50 nm. Figure 5 shows the concept for a monolithic active mirror, where photolithographic techniques are used to define the size and number of actuators. Another simplification of space AO is in the electronics that drive the active mirror. High bandwidth active mirrors require one high voltage amplifier per actuator. PMN/PZT actuators, however, are totally capacitive and require power only when the actuator is moved. No power is required if the mirror is asked to "hold" a particular position. Hence, a single amplifier with a multiplexer can be used to drive the whole active mirror in space.

### 3.3. Integrated monolithic deformable mirror for space

Because of the high actuator densities required, the deformable mirror will require an integrated method of fabrication. One design concept is shown in Fig. 5 (right panel). A single piezoelectric wafer sandwiched between electrodes is used to provide the forces necessary to deform the mirror. Because of the initial quality of the HST optics the active mirror has a low stroke requirement ($\approx +/-50$nm). Consequently, the PZT wafer does not need to be diced. An electrode pattern is deposited on the bottom layer of the PZT wafer. This electrode pattern is used to drive local regions in the PZT wafer. Another metal layer is deposited on the top of the PZT and serves as the ground plane. The electrode pattern will have to be fabricated using photolithographic techniques and will require multiple layers to run signal lines to the periphery of the device. The fabrication procedures must be performed at temperatures which do not depole the PZT wafer and must result in a flat device so print-through effects do not show up on the top surface of the integrated PZT wafer. The whole sandwich assembly is bonded on a glass substrate for structural support. A thin glass surface is then glued on the top of the PZT sandwich. Finally this surface is superpolished and coated to provide an extremely smooth reflecting surface for the mirror.

## 4. SCATTERING: THE *DARK HOLE* ALGORITHM

### 4.1. Introduction

As described in § 2.2., scattering is an important limitation distinct from diffraction. The idea is to use the active mirror to control the scattering level by setting the stroke of each actuator. For example, a perfect wavefront with a small sinusoidal error in the pupil would result in two peaks in the image plane located symmetrically at the corresponding spatial frequency of the sinusoid. The deformable mirror, if it has enough actuators, can remove the sinusoidal error in the pupil and hence the scattered light at the position of the peaks in the image plane.

Given a finite number of actuators, not all the phase errors can be eliminated and there will be some residual scatter. The goal is to reduce the scattering level down to less than $10^{-7}$ of the star intensity at the expected location of the planet. The algorithm that we developed to calculate the actuators stroke is called the **dark hole** algorithm.

The phase at the pupil is then given by:

$$\phi(u,v) = \phi_1(u,v) + \sum_k a_k \sigma_k(u,v), \qquad (5)$$

where $\phi_1(u,v)$ is the figure error of the optics, $\sigma_k(u,v)$ is the influence function from the $k^{\text{th}}$ actuator and $a_k$ is the stroke of the actuator. In our simulations we assumed that the influence functions are only piston only actuators, i.e.

$$\sigma_k(u,v) = \begin{cases} 0 & \text{if outside of sub-pupil } k \\ 1 & \text{if inside of sub-pupil } k \end{cases} \qquad (6)$$

### 4.2. Max Strehl

Usually, active optics are used to maximize the Strehl ratio of the image. This involves minimizing the rms wavefront error over the entire pupil and is equivalent to setting the average wavefront error on each actuator equal to zero. This results in decreasing the scattered light near the center of the image and is shown on Fig. 6 (bottom left panel).

### 4.3. Nonlinear solution

In the case of planet detection we do not necessarily want to suppress scatter near the center of the image but rather at the expected location of the planet (0.5–1 arcsec). In order to determine the actuator stroke $a_k$, we used a chi-square metric consisting of the total intensity in all the dark hole points. We defined a dark hole point, a point of the image where we wish to remove the scattering. The metric was then minimized using a Levenberg-Marquardt nonlinear least-square program. A more detailed description of the nonlinear least-square program is in preparation.[4]

A $1024 \times 1024$ array of points was used to sample both the pupil and the image plane. The pupil was actually only 204 pixels across. Figure 6 shows the result of the dark hole algorithm: it shows the cancellation of the stray light in an annular region with 0.11 arcsec inner radius and 0.82 arcsec outer radius and a square spacing of 0.07 arcsec between the points. A grid of $37 \times 37$ actuators was used with 644 effective actuators. The zero constraint has been applied to 568 image points. Panel (a) shows the location of the dark holes in the image plane. Panel (b) displays the resulting intensity pattern in the image plane and panel (c) shows a radial average of the image intensity. In this case, we achieved a scattering level less than $3 \times 10^{-8}$ below diffraction and a factor of 10 to 100 over the Max strehl solution. The algorithm converged to a solution which had a maximum of $+/-60$nm of stroke. It is important to notice that the scattered light is not lost, but redistributed outside of the dark hole region.

### 4.4. Sensitivity issues

The scattering level that can be achieved by the dark hole algorithm can often be made arbitrarily small by increasing the number of nonlinear least square iterations. In reality, however, the performance of the active optics will be limited by the knowledge of the wavefront aberrations and the surface quality of the deformable mirror. Errors due to imperfect knowledge of the phase errors in the starlight wavefront are modeled as random errors in the actuator positions. We found that we need a 1 nm accuracy in both wavefront sensing and surface quality of the active mirror in order to achieve the results shown above.

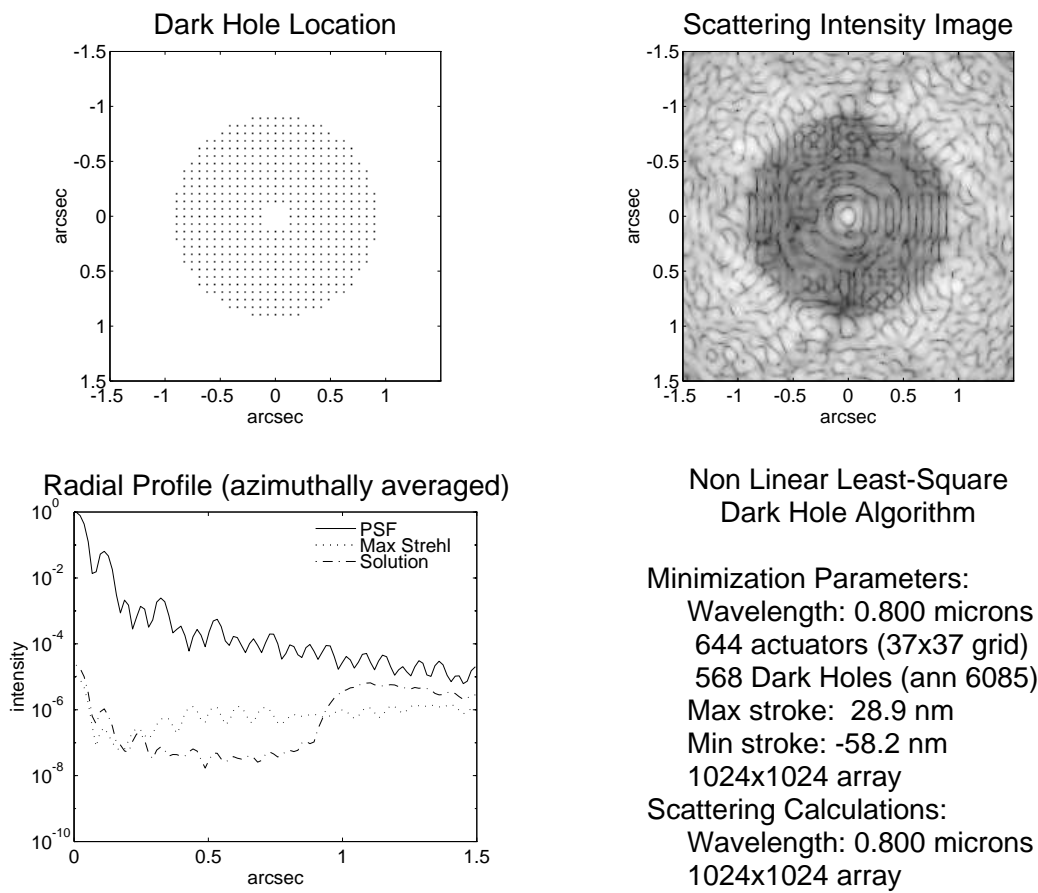

**Figure 6:** Results from the dark hole algorithm. The parameters are shown in the lower right part of the figure. More details are given in the text.

The dark hole algorithm computes the optimum actuator spacing at a particular wavelength. For a fixed actuator configuration, when the wavelength changes two things occur. First the depth of the dark hole changes in proportion to $\lambda^2$. Secondly, the distance of the dark hole from the center of the image plane will change linearly with the wavelength. In order for the dark hole algorithm to function over a broad wavelength band, it is essential that there is significant overlap in the dark hole regions over the entire observation spectral band. Consequently it is important to cancel stray light in regions as close to the center of the field of view as possible.

## 5. DIFFRACTION: CORONOGRAPHY

### 5.1. Introduction

A generic coronograph consists of an occulting mask introduced in the optical beam at the focus of the telescope in order to suppress on-axis image. Lyot[5,6] pointed out that the efficiency of a coronograph was appreciably increased by using a Lyot stop, slightly smaller than the telescope aperture in a pupil plane following the focal mask (see Fig. 7).

Basically, a coronograph is an optical cascade. The first occulting mask has two purposes: (1) to reduce the light coming from the central star by a direct occultation, and (2) to filter the low spatial frequency of the pupil. This results in: (1) a reduction of the stellar light in the pupil plane, and (2) a redistribution of the remaining light to the edge of the pupil. The purpose of the Lyot stop is then to mask the light located on the edge of the pupil. For a point located away from the central star, the occulting mask has a negligible effect and the Lyot stop decreases only the size of the aperture causing a slight increase in the width of the off-axis PSF.

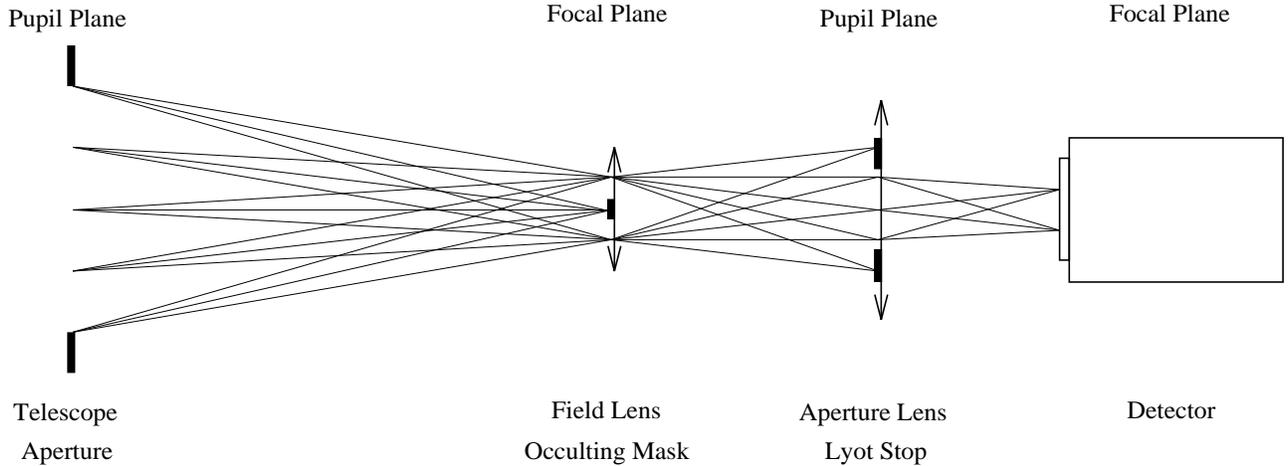

**Figure 7:** Optical lay-out of a Lyot coronograph

### 5.2. Choices of the occulting mask and of the Lyot Stop

In order to get the best star light rejection, we need to optimize the transmission profile of the occulting mask and the Lyot stop. The simplest shape is called sharp-edge and is a stop which has a transmission of 100% out of the mask and 0% inside the mask. This kind of mask is easy to build, but it is not very efficient, because the sharp edge creates diffraction rings in the pupil.

The amplitude in the Lyot plane is the convolution of the input pupil (the occulting mask is stopping all the central stellar light but the outer rings which are located at the pupil edge) by the Fourier transform of the occulting mask shape. If the occulting mask is gaussian, the best shape for the Lyot stop is the input pupil shape with a gaussian transition to the edge. The width of this transition is inversely proportional to the size of the focal mask.

However we chose another approach which maximizes the signal to noise ratio in the image plane and therefore in the Lyot pupil plane. The best way of accomplishing this is to compute the intensity of a perfect on-axis source in the Lyot plane and then to use a mask which is the inverse of this intensity distribution. In order to place an upper limit on the mask we use a threshold level. The part of the pupil which is below the threshold level has a 100% transmission and the rest has a transmission inversely proportional to the intensity of this perfect source. In order to further increase the slope of the brightness radial distribution, we multiplied this filter by a filter having the following radial transmission:

$$T(u,v) = \left[1 - \left[\frac{\rho - (r_1 + r_2)/2}{(r1 - r2)/2}\right]^2\right]^2, \qquad (7)$$

with $r_1$ the pupil radius, $r_2$ the central obscuration radius.

### 5.3. Simulations

A $1024 \times 1024$ array of points was used to sample both the pupil and the image plane. We took a perfect wavefront with the shape of the actual HST pupil. Figure 8 shows the result of the coronograph simulation. The gaussian mask has a FWHM corresponding to four Airy rings. The transmission is 88% at 0.5 arcsec from its center. The Lyot stop transmits 21% of the input pupil area resulting in a total throughput for the planet of 18%.

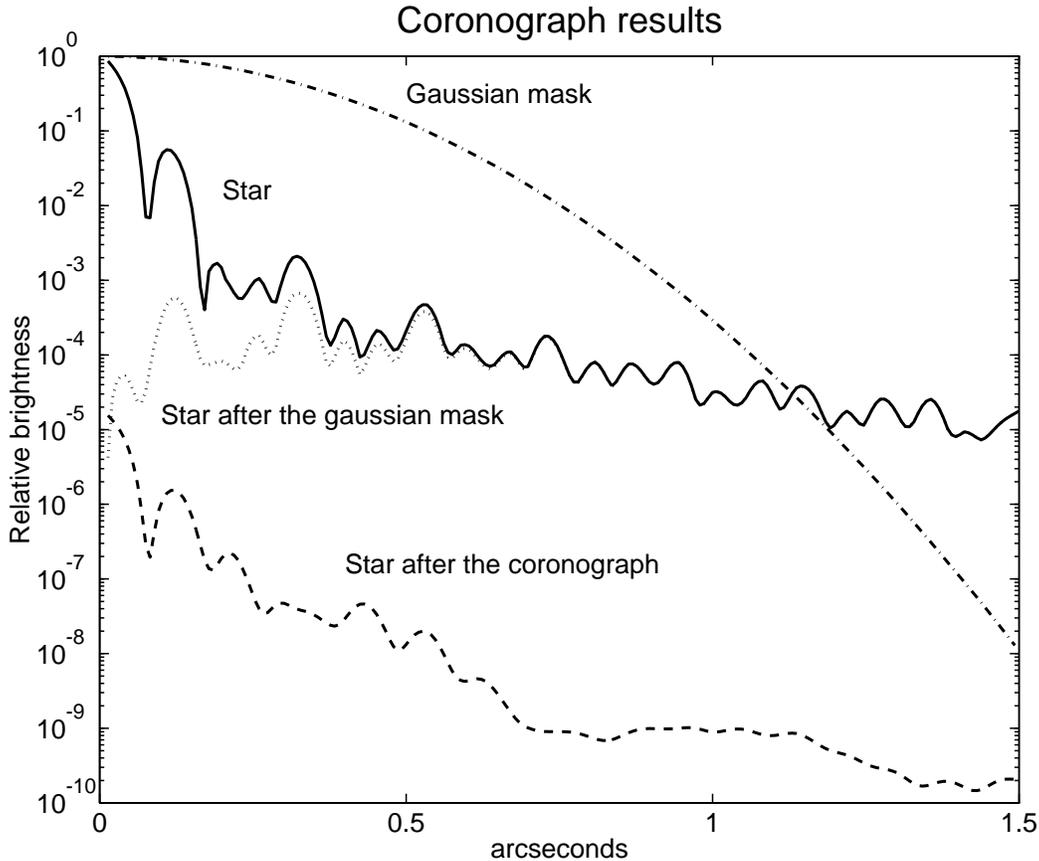

**Figure 8:** Results from the coronograph simulations. See text for more details.

## 6. THE SPACE TELESCOPE ADVANCED CAMERA AO/CORONOGRAPH

The goal of this study is to design an Active Optics Coronograph for the Hubble Space Telescope in order to image extra-solar planets directly. The instrument would be composed of optics to re-image the pupil on a deformable mirror. The spherical aberration of HST would be corrected by designing the active mirror with a compensating curvature. The focal occulting mask would be located behind and followed by a Lyot stop at a reimage of the pupil. A fold mirror would articulate in order to align the active mirror with the primary.

To estimate the actual exposure needed to detect a planet, we used the transmission given in Table 2. The number of photons received per second from the planet is then 0.2 photon/s and requires 1h4m of integration time to achieve a SNR = 5 (cf. Table 3).

## 7. CONCLUSION

We have proposed a technique which will enable the Hubble Space Telescope to detect extra-solar planets in a reasonable amount of exposure time. The technique uses space active optics with a large number of actuators in order to reduce the wavefront errors of the primary and secondary mirrors, and a coronograph to reduce the amount of diffracted light from the central source. In the future we will be demonstrating this technique in the laboratory.

## 8. ACKNOWLEDGEMENTS

We would like to thank Claude Roddier for providing the Hubble Space Telescope wavefront data. We would like also to thank R. Terrile, S. Pravdo and C. Ftaclas for stimulating discussions. This work was performed at

Table 2: Exposure time parameters

| Hubble Space Telescope | | Active Optics/Coronograph | | | | Detector | |
|---|---|---|---|---|---|---|---|
| Primary mirror | 87% | Deformable mirror | 98% | Focal mask | 88% | Optics | 90% |
| Secondary mirror | 87% | Relay lenses | 90% | Lyot stop | 21% | QE | 80% |
| Transmission | 75% | Transmission | 17% | | | Transmission | 72% |

| Total throughput for the planet = 9% |
|---|
| Central wavelength = $0.8\mu$m |
| Bandwidth = $0.4\mu$m |

| Astrophysical parameters | | | |
|---|---|---|---|
| Star distance | 10 pc | Planet orbit | 5.2 AU |
| Star radius | $1R_\odot$ | Planet albedo | 0.5 |
| Star temperature | 5770K | Planet radius | $7\times 10^7$m |

Table 3: Exposure time result

| Number of photons detected per second: | |
|---|---|
| star | $2\times 10^8$ |
| background | 5.2 |
| planet | 0.2 |
| Exposure time required | |
| Signal/Noise Ratio | 5 |
| integration time | 1h 4m |